# Green open access in computer science – an exploratory study on author-based self-archiving awareness, practice, and inhibitors

Daniel Graziotin*

Faculty of Computer Science, Free University of Bozen-Bolzano, Bolzano, Italy
*Corresponding author's e-mail address: daniel.graziotin@unibz.it

## ABSTRACT

Access to the work of others is something that is too often taken for granted, yet problematic and difficult to be obtained unless someone pays for it. Green and gold open access are claimed to be a solution to this problem. While open access is gaining momentum in many fields, there is a limited and seasoned knowledge about self-archiving in computer science. In particular, there is an inadequate understanding of author-based self-archiving awareness, practice, and inhibitors. This article reports an exploratory study of the awareness of self-archiving, the practice of self-archiving, and the inhibitors of self-archiving among authors in an Italian computer science faculty. Forty-nine individuals among interns, PhD students, researchers, and professors were recruited in a questionnaire (response rate of 72.8%). The quantitative and qualitative responses suggested that there is still work needed in terms of advocating green open access to computer science authors who seldom self-archive and when they do, they often infringe the copyright transfer agreements (CTAs) of the publishers. In addition, tools from the open-source community are needed to facilitate author-based self-archiving, which should comprise of an automatic check of the CTAs. The study identified nine factors inhibiting the act of self-archiving among computer scientists. As a first step, this study proposes several propositions regarding author-based self-archiving in computer science that can be further investigated. Recommendations to foster self-archiving in computer science, based on the results, are provided.

## INTRODUCTION

While there is a never-ending debate regarding what constitutes science and the progression of scientific advancement [1–3], it is difficult to argue against the claim that science relies on the availability of knowledge. Knowledge is constructed by individuals during activities that are sometimes in cooperation, sometimes in competition, but always in the context of communities [4]. Researcher activities rely on data collection, analysis, publication, and the critique and reuse of someone's work [5]. Therefore, access to the work of others is necessary in order to evaluate, replicate, and build upon that knowledge [6].

Unfortunately, access to the work of others is something that is too often taken as granted. Unless paid by someone, access to this work is both problematic and difficult to acquire. Knowledge has been buried systematically throughout history; first behind the walls of thousands of geographically sparse libraries and then behind the *paywalls* of digital systems [7].

Nowadays, scholarly publishers are playing an essential role in knowledge sharing between research parties. They guard the majority of published academic knowledge. Thus, publishers are still essential for the efficiency of research and the dissemination of knowledge [8]. However, publishers are often for-profit entities with large margins [9]. These margins are achieved by erecting expensive paywalls between the published knowledge and the readers [10]. Paywalls cause knowledge to be inaccessible to many researchers. Universities and research centers can only afford a small portion of subscription-based journals [11] because the subscription costs are rising [12] even faster than inflation [13].

This situation has been said to change soon [14, 15]. It has even been argued that one day traditional academic journals will not exist anymore [16, 17]. Meanwhile, paywalls cause research opportunities to become inevitably lost, and the quality of the research performed to be in danger.

Marking all research articles, data, and general artifacts freely available on the Internet is a solution to this problem. Open access is one such initiative [18–20]. Different definitions for open access exist as well as different publishing models. These definitions have been discussed extensively elsewhere [21]. Regarding the publishing models, in short, the publishers joining the *golden road* will have all their journal articles freely accessible under a license supporting open access, which is often one of the Creative Commons licenses. Publishers taking the *green road* allow researchers to self-archive their own generated version of scientific articles in their personal website or in repositories of publications [11]. Subscription-based journals may adopt a *hybrid* model between the golden road and the traditional model, where authors pay fees in order to open the electronic versions of articles on an individual basis [22].

Green open access is the subject of this article and thus must be defined properly. Green open access publishers are the traditional publishers—those of subscription-based journals— that allow researchers to self-archive pre-publication versions of papers. The particular versions of the author-generated articles, and the venues where these papers can be archived, vary strongly among publishers. In other words, green open access (or self-archiving) happens in different combinations of the following three levels: what is self-archived, where self-archiving happens, and when self-archiving happens.

Regarding what is self-archived, it is challenging to define clearly the stages of a research paper. For example, an article is written collaboratively by one or more scientists and then submitted to an academic journal. At that point, before the start of the process of peer review, the article is commonly called a *preprint*. After the manuscript is accepted for publication, the copyright is transferred to the publisher who performs typesetting and proofreading. Paywalled or not, the scientific article, at that point, becomes available to the general public via digital libraries, typically in PDF format. This is commonly called the *publisher's article* (or publisher's PDF). This leaves a gap regarding anything in between. After submission and before publishing, the article faces one or more round of reviews. Consequently, the authors address the reviewers' and editors' comments at each revision. Inevitably, the article's contents change between submission and publishing. Some publishers call *postprint* each author-created version of a manuscript before the typesetting and publishing services of the publishers themselves. Other publishers instead define preprint as any revision of a manuscript created during the peer review process and postprint as the one which is accepted for publication.

Regarding where self-archiving happens, three possible types of venue have been identified: in a personal or academic website (including one of a research group), in a digital preservation repository of an institution, or in general public repositories of publications that are sometimes multidisciplinary [11, 23]. Examples of the latter include arXiv, figshare, zotero, the Social Science Research Report (SSRN), Research Papers in Economics (RePEc), and PeerJ PrePrints.

Regarding when self-archiving happens, traditional publishers can impose a delay before pre- and postprints can be self-archived. Some major publishers are going into this direction, allowing self-archiving only 12 months after publication [24]. This rule sometimes applies only to repositories, while self-archiving of preprints might still be allowed on personal and academic websites. As it has been discovered that only 34% of URLs remain operational after a four-year period [25], it is not really surprising to see publishers keep letting authors to self-archive on their personal websites immediately after publication. Arguably, many authors would either forget or avoid self-archiving in repositories after 12 months. Persisting self-archiving is ensured by established archived or distributed repositories like those mentioned in the previous paragraph.

The copyright transfer agreements (CTAs) between the publishers and the authors regulate what and where self-archiving happens. However, it appears that CTAs are difficult to be understood or ignored by authors. Despite the fact that that 90% of publishers allow self-archiving, only less than 20% of the papers have been self-archived [11].

In another recently published article [21], the golden open access journals in computer science (in particular, in software engineering and information systems) were analyzed. The results of the systematic analysis pictured an obscure panorama. The majority of the journals were unknown, lacked transparency, did not offer the archival of articles, shaded their review process and publication ethics, and asked for too large article processing charges that were completely unjustified by the features offered. Consequently, gold open access in computer science is still in its infancy. When mentioning open access to researchers and professors in computer science, their reactions are often lacking understanding or are biased by distrust. More work should be done on the publisher side and when advocating gold open access in computer science. On the other hand, the willingness of subscription-based publishers to allow authors to self-archive pre-publication versions of their articles is increasing [11]. The most important publishers in computer science allow author-based archiving of articles [21]; thus, present authors may wonder about the state of green open access in computer science as well[1].

Less than 20% of published articles have been self-archived [11, 26]. Rarely, articles in institutional repositories have been archived by their own authors [27]. However, the status of green open access in computer science is mostly unknown, although some claim that computer science is one of the long-standing self-archiving communities [28]. It appears that there are a limited number of studies regarding green open access and computer science. Lawrence [29] analyzed around 120,000 formal publication papers presented at computer science conferences. The self-archived papers in the sample were cited 157% more than those non-freely available. The increase, in terms of citations, went up to 286% on average when considering the top-tier conferences. The percentage of self-archived papers in the sample was not disclosed in the paper. Similar results were achieved in other studies analyzing different fields [30, 31]. It should also be noted that in other disciplines, such as astronomy, this effect could not be observed [32].

---

[1]. This exploratory study was conducted with seriousness and scientific objectivity into mind. However, the present author feels obliged to disclose his sympathy for open access and open science in general. The author *believes* that opening science has more advantages than disadvantages, the published articles never suffered from this. For example, Graziotin et al. [21] criticize open access journals in the fields of software engineering and information systems with the aim to improve the perception of open access in these fields.

Lawrence's study has been considered the first regarding green open access [33]. The technical report by Swan & Brown [28] claimed that computer science is the discipline most prone to self-archiving. The authors stated that "Today, there are more articles […] freely available through self-archiving in computer science than in any other subject" ([28], p. 1). The authors found that computer science has been a leading discipline with respect to self-archiving. However, the report is not a peer reviewed study, and it will soon be 10 years old. Therefore, the understanding provided by the report is outdated and it might not be considered suitable for academic consideration. Miller [34] conducted a survey with a sample of 443 UK individuals of which around 85% declared to work in the field named "computing." Less than 10% of the participants in the computing field were not aware of self-archiving. Between 70% and 95% of the computing participants declared to employ self-archived papers in their research activities. The thesis analyzed attitudes toward self-archived articles but did not analyze the inhibitors of self-archiving. There is still a need to understand the status of green open access in computer science, especially with respect to the factors inhibiting self-archiving.

The present author's experience when mentioning self-archiving to colleagues, visiting academics, and authors at conferences has been miserable. The majority of the informal talks either highlighted a lack of knowledge of self-archiving allowance or reported the unknowingly illegal practice of hosting the publishers' PDF in personal websites. Academics showed a lack of understanding about the rights kept or given away when signing the CTAs. Some of them showed a lack of basic understanding of what copyright is. Authors were often unaware of the fact that self-archived papers are significantly more cited than paywalled articles [26, 29]. Most of the encountered authors were not even aware of the existence of arXiv, which is the de facto standard multidisciplinary repository for computer science fields [26].

This article presents an exploration on author-based self-archiving awareness, practice, and inhibitors in computer science. It is an objective of this study to provide an initial understanding of the phenomenon. The results of this study are limited by its exploratory nature. However, the results offer several propositions grounded in the evidence. These propositions should be further explored in future studies.

## MATERIALS AND METHODS

The research methodology adopted was the survey. A Web-based questionnaire was developed and distributed through the internal mailing list of the faculty of computer science at the Free University of Bozen-Bolzano.

The questionnaire was limited to eight questions in order to cope with the limited amount of time that academics usually have. Two of the questions were for demographic purposes, making the core of the questionnaire six questions.

The questionnaire started with a short introductory text, which thanked the participants in advance for participation and promised the participants that the questionnaire would require only five minutes of their time.

Question 1 was: *What role describes you best?* It was pre-populated with the answers "PhD student," "Researcher," and "Professor." There was a fourth open-ended option named "Other (please specify)."

Question 2 was: *Approximately how many peer-reviewed, published papers are you listed as an author of?* The possible answers for this question were "Less than 20," "between 20 and 50," and "More than 50."

Both the first and second questions were for demographic purposes only. Other possible questions could have been inserted such as the age, sex, and country of provenance of the participant. They were not seen as questions of value given the added time needed to complete the questionnaire.

The real questionnaire started at Question 3: *Do you know what the term "self-archiving" means?* The answer was of a binary type, "yes" or "no." Publishers often employ the phrase self-archiving (or selfarchiving) rights in CTAs. Thus, it is important for authors to understand the actual meaning of the term.

Next was Question 4: *Can you clearly differentiate the terms "eprint," "preprint," "postprint," and "publisher PDF?"* While *eprint* is not a widely employed term in CTAs, it is employed in established venues like arXiv and has been used in prior studies. On the other hand, the terms *preprint* and *postprint* appear on the CTAs of major computer science publishers. It is important to understand whether authors are able to differentiate the terms. The answer was of a binary type, "yes" or "no."

Question 5 was preceded by a short text explaining the meaning of the terms self-archiving, e-print, preprint, postprint, publisher's PDF, and where self-archiving happens. The reason for these explanations was to level the knowledge of the participants. The remaining items of the questionnaire assumed knowledge of these terms. The following was the text presented by the questionnaire:

> Here are some informal definitions: Self-archiving is the act of depositing an e-print on the World Wide Web, so that it is accessible without any barrier.
>
> An e-print is a digital version of a research document (a PDF, a .docx, an .odt, etc.), which can be in the state of preprint, postprint, and the publisher PDF.

A preprint is a digital version of a research document before it is submitted for peer-review. A preprint can be a research proposal, a draft of a scientific article, but also a manuscript that it is submitted after major or minor revisions are required.

A postprint is a digital version of a research document that it has been accepted for publication. It is generated by the authors of the article.

The publisher PDF is the research article as it can be downloaded from the journal/proceedings/publisher website, e.g., ACM Digital Library and IEEE Xplore.

Self-archiving can happen at three possible levels: (1) personal (or academic) website, (2) institutional repository, (3) (multi)disciplinary repository

Question 5 was needed to assess the knowledge of the participants regarding the possibility to self-archive. The answer type was binary, "yes" or "no."

> Did you know that the major publishers in Computer science-ACM, IEEE, INFORMS, Elsevier, ME Sharpe, Palgrave Macmillan, Springer Verlag, John Wiley and Sons-allow you to self-archive at least the preprints of a research article?

The two upcoming questions assessed the frequency of self-archiving and the preferences for hosting the e-prints.

Question 6 was: *With respect to your previous publications, how often have you self-archived?* It comprised of three items completing the question: *a preprint (including drafts)*, *a postprint*, and *the publisher's PDF*. The answers were Likert items, in the range *Never, Very Rarely, Rarely, Occasionally, Very Frequently*, and *Always*.

Question 7 was: *With respect to your previous publications, where have you self-archived?* It comprised of three items completing the question: *personal (or academic)* website, *an institutional repository*, and *a (multi)disciplinary repository*. The answers were Likert items, in the range *Never, Very Rarely, Rarely, Occasionally, Very Frequently*, and *Always*.

Question 8 was open-ended and non-mandatory. It asked: *What prevents you to self-archive your scientific articles?* The answer could be of any length.

## RESULTS

The questionnaire was administered in November 2013. It remained open for three days. The invitation to participate was solicited via the internal mailing lists of the Faculty of Computer Science at the Free University of Bozen-Bolzano. In addition, the present author personally invited faculty members by constructing businesscard-sized reminders containing a human-readable shortened URL of the questionnaire.

Although it was not possible to obtain the number of people tied to the internal mailing list, the faculty comprised of 66 members according to its website. Forty-nine of them participated in the questionnaire. However, one respondent did not reach the questions after those of demographic nature. The response rate was 72.8%. Of those responding to the questions, 39.6% were researchers, 37.5% were PhD students, 16.7% were professors, and 6.2% were interns.

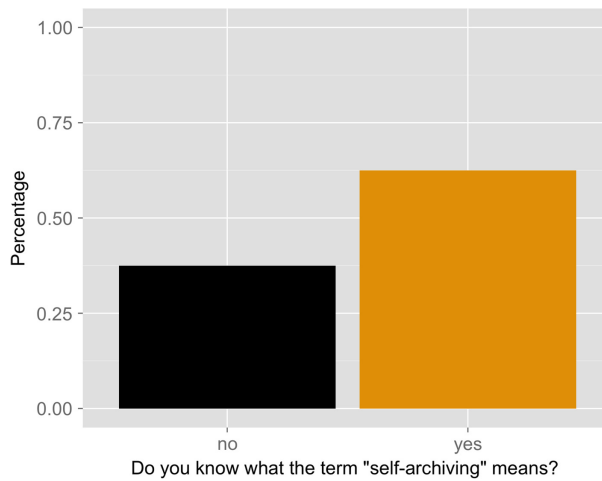

**Figure 1.** Answers to the question: *Do you know what the term "self-archiving" means?*

The majority (67.4%) of the participants published less than 20 papers at the time of the questionnaire; 12.2% of the participants published between 21 and 50 papers, while 20.4% of them were an author of more than 50 scientific articles.

For the question: *Do you know what the term "self-archiving" means?* 62.5% of the participants answered positively as illustrated by Figure 1.

Almost the opposite result happened when responding to the question: *Can you clearly differentiate the terms "eprint," "preprint," "postprint," and "publisher's PDF"?* Figure 2 shows that more than half of the participants (64.6%) were not able to differentiate between the meanings of these terms.

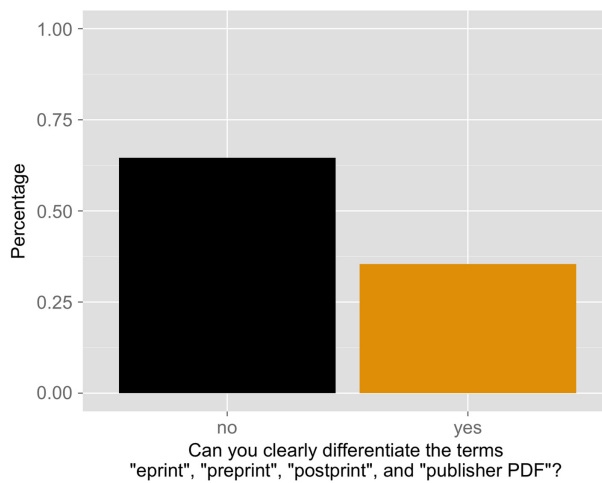

**Figure 2.** Answers to the question: *Can you clearly differentiate the terms "eprint," "preprint," "postprint," and "publisher PDF"?*

After this question, the participants read the definitions related to self-archiving which were detailed in the previous section. Their response to the question, *Did you know that the major publishers in computer science allow you to self-archive at least the preprints of a research article?*, was almost half-and-half; 54.5% of the participants were aware of self-archiving allowance by the publishers. Figure 3 illustrates the answers to the question.

Figure 4 summarizes the responses to the question: *With respect to your previous publications, how often have you self-archived a preprint, a postprint, and the publisher's PDF?* As it can be seen, at least half of the participants never self-archived any paper in any possible form. The results achieved in this question are consistent with those of the previous question.

Figure 5 summarizes the responses to the question: *With respect to your previous publications, where have you*

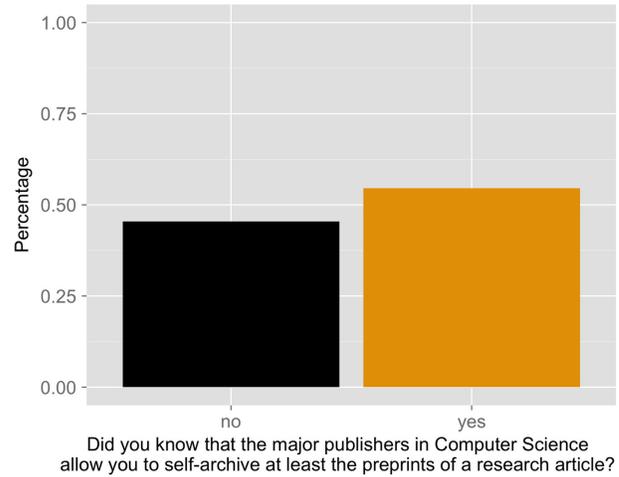

**Figure 3.** Answers to the question: *Did you know that the major publishers in computer science allow you to self-archive at least the preprints of a research article?*

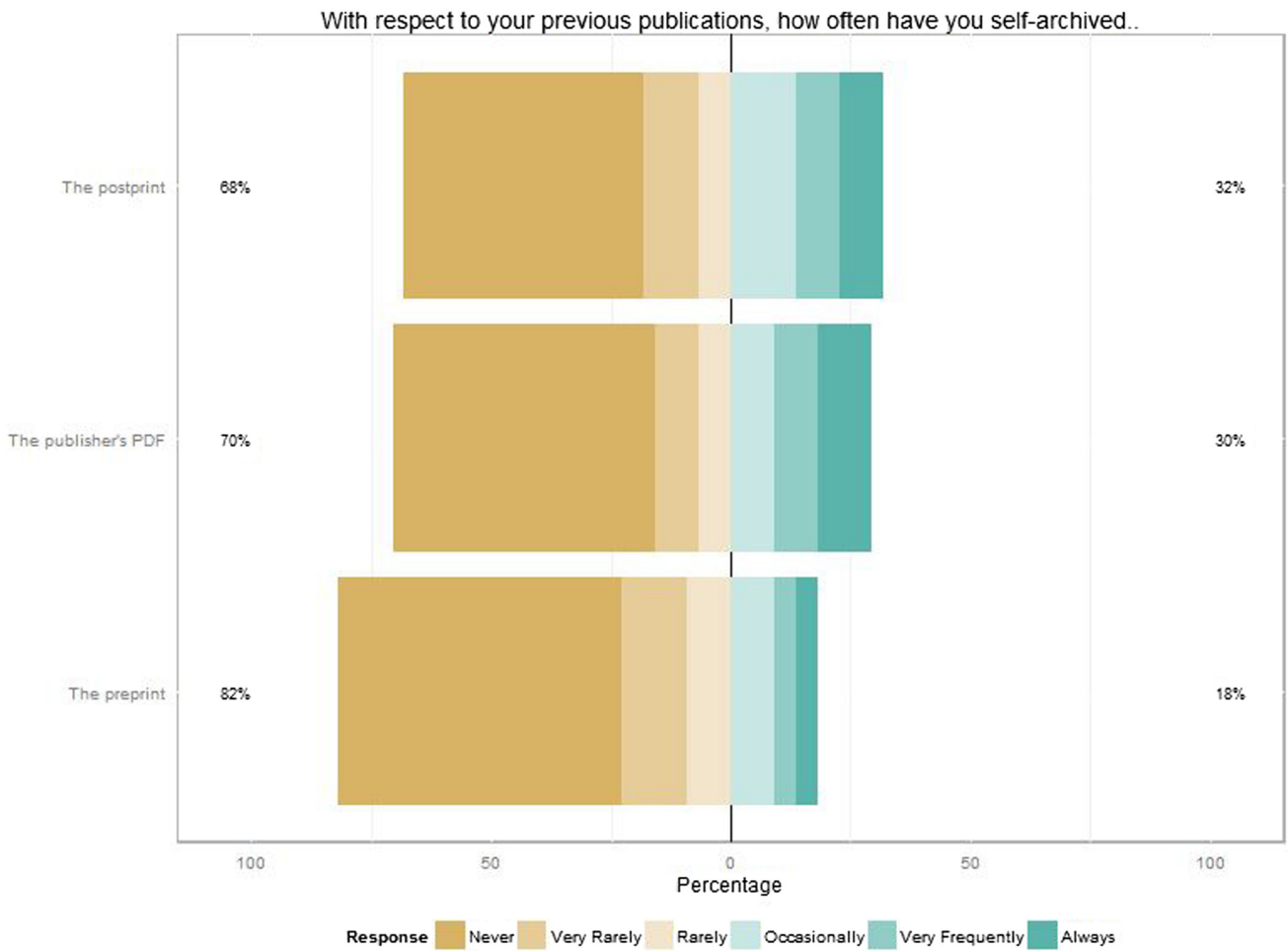

**Figure 4.** Answers to the question: *With respect to your previous publications, how often have you self-archived a preprint, a postprint, and the publisher's PDF?*

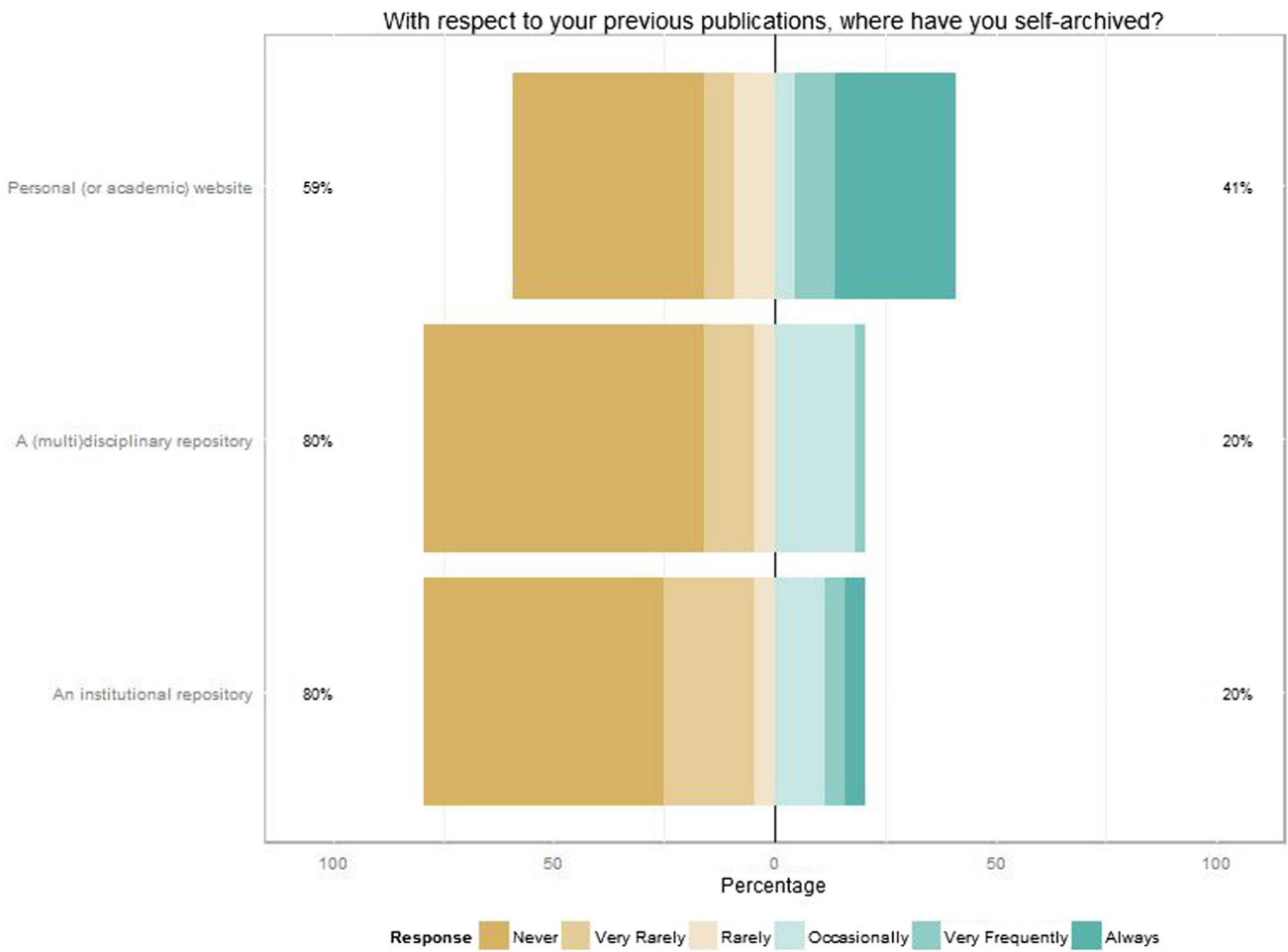

**Figure 5.** Answers to the question: *With respect to your previous publications, where have you self-archived?*

*self-archived?* About half of the participants never self-archived any paper in any possible venue; 63.6% of them never employed a multidisciplinary repository like the arXiv. Among the respondents who declared to not be aware of the allowance of self-archiving in the CTAs, 25% of them very rarely to occasionally self-archived preprints; 45% of them very rarely to always self-archived postprints; and 45% of them very rarely to always self-archived the publisher's PDF.

The open-ended, non-mandatory question, *What prevents you to self-archive your scientific articles?*, was answered by 72.9% of the participants. Figure 6 is a word cloud[2] of the 100 most frequent words provided by the respondents.

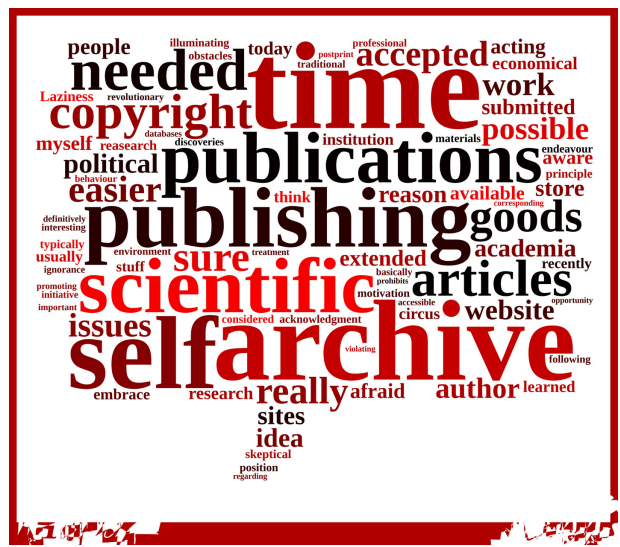

**Figure 6.** A word cloud of the 100 most frequent words provided by the respondents.

_______________

2. Generated by http://www.tagxedo.com, changed options: theme "Black meets Red" 000000 330000 660000 990000 CC0000 FF0000; Font: liberation Serif; Orientation: horizontal; Shape: callout; Maximum word count: 100; Word manually removed: "don"

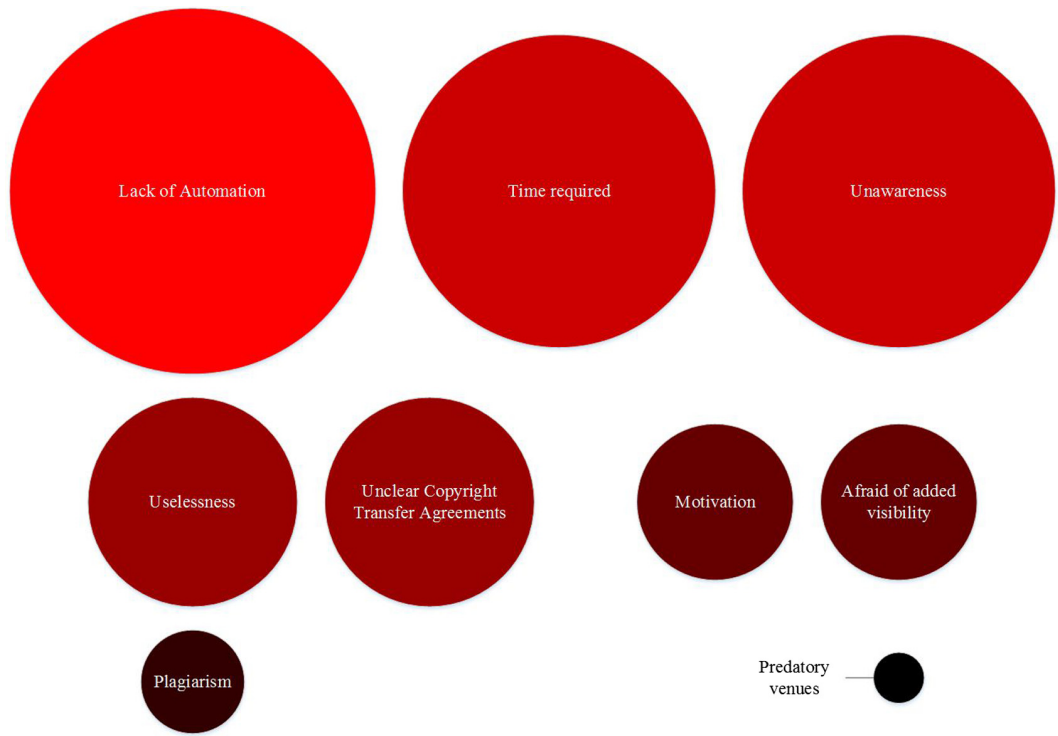

**Figure 7.** Identified factors inhibiting self-archiving.

A first model of the factors inhibiting self-archiving in computer science emerged from *in vivo* and open coding of the open-ended responses. The nine factors are represented in Figure 7. The factors are colored bubbles. The diameter of the bubbles is proportional to the frequency of occurrences of the nodes in the open-ended answers. The background color of the bubbles is a gradient from red to black, as suggested by Fronza et al. [35], to further guide the reader in identifying the magnitude of the factors.

As illustrated in Figure 7, the data analysis could identify nine factors. The factors, given in order of importance in terms of frequency, are lack of automation, time required, unawareness, uselessness, unclear CTAs, motivation, afraid of added visibility, plagiarism, and predatory venues.

The lack of automation factor was elicited by data such as "[I do] not know about good tools to make it [self-archiving] easier for me." This factor appears to be much related to the second biggest factor, the time required. The time required factor was coded *in vivo* from responses like "time" and "the time required for completing the task." Unawareness of self-archiving is the third most important factor identified. It was elicited from sentences like, "I did not know about it [self-archiving] before." The fourth most frequent code was a perceived uselessness of self-archiving. The comments on this aspect spaced from informed belief of uselessness, e.g., "Preprints will be changed anyway, so no point in publishing them" to comments soaked with misinformation of the academic publishing system, e.g., "Why should I do it? Publications are accessible via other databases, so I don't see the need to do it." Authors found CTAs to be difficult to comprehend as "the term imposed by the publisher must be read carefully." Interestingly, one participant found it difficult to understand whether self-archiving is possible "[…] regarding papers of which I am only a co-author, but not the main or corresponding author." Some comments reported avoiding self-archiving because of the unclear CTAs and a real fear of the publisher's possible reactions. Several participants felt unmotivated to self-archive, citing "nothing but myself get started" and "laziness" as the main sources of lack of motivation. This is further evidence that advocacy of self-archiving in computer science is needed. Participants seemed to be concerned with the added visibility that self-archiving would bring them. For example, a researcher was afraid of "not being able to keep track where I give out and store my publication." Some participants were also "not very eager to have some of my papers available on my website." Afraid of plagiarism is another factor preventing self-archiving. Some participants were afraid that "someone else can steal [my] personal work." These participants were probably not aware that work can be stolen regardless of it being the publisher's PDF or a self-archived preprint. Finally, one participant was actually afraid of "fraud actions" when self-archiving, allegedly leading to predatory venues for self-archiving.

## DISCUSSION

The results obtained from the questionnaire enable the formulation of propositions regarding the status of self-archiving among computer scientists to be addressed in future studies. The purpose of this study is to offer expected magnitudes of the results of future studies rather than generalizing to precise quantifications. Therefore, to further highlight the exploratory nature of this study, the percentage stated in the following propositions has been rounded to the nearest multiple of 5.

- **Proposition 1.** We expect that 60% of the computer scientists know what the term self-archiving means.

Proposition 1 entails that we expect that 40% of researchers in computer science do not know what the term self-archiving means. While not knowing the term self-archiving does not necessarily mean that scientists are not performing it, self-archiving is the term most employed in CTAs. It is necessary for academics to understand this term.

Despite the fact that 60% of the respondents knew the meaning of the term self-archiving, the fourth question showed that roughly the same amount of scientists are not able to differentiate the terms preprint, postprint, and publisher's PDF.

- **Proposition 2.** We expect that 60% of the computer scientists cannot differentiate between eprint, preprint, postprint, and publisher's PDF.

Proposition 2 suggests that both the CTAs and the general terms employed by advocates of open access are not clear enough. It is advisable to advocate open access more clearly in the field of computer science.

The responses to the fifth question suggested the following proposition regarding the knowledge of self-archiving rights.

- **Proposition 3.** We expect that 55% of computer scientists are aware that the major publishers in the field self-archive at least the preprints of scientific articles.

Proposition 3 indicates that misunderstanding and lack of knowledge are seriously impacting self-archiving in computer science. About half of the researchers in this field might not even know that self-archiving is possible and legal.

From the sixth question, we formulate the following propositions about the frequency of self-archiving among computer scientists.

- **Proposition 4.** We expect that 60% of computer scientists never self-archive preprints.
  - 4.1. We expect that 80% of computer scientists never or rarely self-archive preprints.
- **Proposition 5.** We expect that 50% of computer scientists never self-archive postprints.
  - 5.1. We expect that 70% of computer scientists never or rarely self-archive postprints.

Propositions 4 and 5 indicate a serious threat in computer science. Knowledge produced in this subject is at risk to be hindered behind a paywall. Another interesting proposition is offered regarding self-archiving the publisher's PDF.

- **Proposition 6.** We expect that 55% of computer scientists never self-archive the publisher's PDF.
  - 6.1. We expect that 45% of computer scientists very rarely to always self-archive the publisher's PDF.
  - 6.2. We expect that 20% of computer scientists very frequently or always self-archive the publisher's PDF.

Proposition 6 entails that 55% of computer scientists do not self-archive the publisher's PDF. Thus, the CTAs are respected by about half of the participants. However, a consequence of Proposition 6 is that 45% of the researchers in computer science, at various levels of frequency, do not respect CTAs (6.1). On top of that, Proposition 6.2 indicates that 20% of academics in computer science very frequently self-archive the publisher's PDF, breaking the CTAs. Whether they are aware of breaking the rules or not, it is no longer astonishing how surprised several researchers appear when for-profit publishers send takedown notices to academic website owners or services like academia.edu.[3]

From the seventh question, the following propositions are formulated regarding the venues for self-archiving.

- **Proposition 7.** We expect that 45% of computer scientists never self-archive in personal (or academic) websites.
  - 7.1. We expect that 40% of computer scientists occasionally to always self-archive in personal (or academic) websites.
- **Proposition 8.** We expect that 55% of computer scientists never self-archive in institutional repositories.
  - 8.1. We expect that 30% of computer scientists occasionally to always self-archive in institutional repositories.
- **Proposition 9.** We expect that 65% of computer scientists never self-archive in (multi)disciplinary repositories.
  - 9.1. We expect that 20% of computer scientists occasionally to frequently self-archive in (multi) disciplinary repositories.

From the responses to Question 8, nine inhibitors of self-archiving were identified. Those factors are expected to sit in different layers, in terms of severity and importance, among computer scientists. The following propositions are offered.

- **Proposition 10.** We expect that the major inhibitors of self-archiving in computer science are lack of automation mechanisms and tools, the time required for self-archiving, and the unawareness of the self-archiving practice itself.

---

3. Example of comments include those left on academic newspapers like the Chronicle of Higher Education [38], on blog posts of open access advocates [39], and in the social media [40, 41].

- **10.1.** We expect that the intermediate factors inhibiting self-archiving are the perceived uselessness of self-archiving, difficulty to understand and comprehend CTAs, a lack of motivation, and being afraid of the added visibility when self-archiving.
- **10.2.** We expect that the minor factors inhibiting self-archiving lie in a fear of plagiarism, and the belief that there are predatory venues for green open access.

As a consequence of Proposition 10, the recommendation offered by this author is to build software that automatically lets authors self-archive their scientific articles. These tools should minimize the effort required by the researchers. One example could be the automatic generation of an HTML page listing and the subsequent linking of the preprints residing in a folder for uploading to a server on the Web. Another tool would be desirable for automatically posting preprints on arXiv and similar websites. Currently, redundant and easily extractable information such as the title of the paper, the authors and affiliations, and the abstract have to be manually entered when submitting preprints to (multi)disciplinary repositories. The metadata could also easily be gathered from bibliographic services of the published papers, e.g., Mendeley or exported data from Google Scholar.

As a consequence of Proposition 10.1, tools to further simplify the understanding of CTAs are desirable. One tool, SHERPA/RoMEO, is established but lacks modern interfaces, user-friendliness, and immediate display of the relevant information. The Web tool rchive.it is a further step toward this.

Another consequence of Proposition 10, and its sub-propositions, is that more advocacy of open access is required in the fields of computer science. Green and gold open access practices are establishing in the fields of earth science, mathematics, and physics [36, 37]. However, open access, in general, seems to not be understood in computer sciences both in the golden road, as showed in another study [21], and in the green road, as showed in the current article. This is surprising because computer science is the natural field of open source, a very similar phenomenon.

A researcher who published between 20 and 50 papers preceded the answer to Question 8 with the following text. The present author decided to reproduce the text in full because it delivers a profound message that should cause reflections for both advocates and opponents of open access publishing.

we (in the academia) are all well aware that the act of publishing some scientific result is today not only a matter of keeping informed the research community, but rather a political and economical endeavour. It's "political and economical" in the sense that scientific publications are often not considered at all as the physical carriers of new scientific knowledge and illuminating discoveries but as goods that are placed on the market in order to be bought and sold. The more of these goods you are able to sell, the more you can count on funds for your lab/institution and research activity, as well as on new personal career possibility. 99% of the times, we learn how to grow in this environment by following the habit of people older than us, who have invented this circus. they teach us that publishing in the'traditional is the only mean you have to paved the way of your professional success.; this is to say, basically, that I think that self-archiving is a very interesting initiative, and political behaviour. to be fair, nothing prevents me to embrace this way of doing except for my ignorance, and the fact that I usually don't take my publications as something that deserves a so special, and revolutionary, treatment; I think that I will be ready, after all, to fully embrace it (and actively promoting it), as soon as I will have accepted the idea of definitively leaving the scientific circus-market we are living in today, and I will have in my hands some good idea to communicate.

While this study is limited by its exploratory nature, it offers a preliminary set of propositions to be explored in future studies. However, the preliminary evidence hints clearly that there is still a long way before self-archiving can be declared a standard practice among computer scientists. Advocates and software developers are needed in order to sweep away the misinformation, the unawareness, and the laziness, and to automatize the majority of the process of self-archiving.

### ACKNOWLEDGMENTS

This author would like to thank Pekka Abrahamsson and Xiaofeng Wang for their support in both performing open access and studying it as a phenomenon. The present author is grateful to all SFScon 2013 participants and organizers, in particular Patrick Ohnewein. The author is thankful to Science Open for waiving the processing charges of this article. Finally, the author is thankful to eSurvey Creator for providing their full service for free to students.

### REFERENCES

[1]   Feyerabend P. Against method [Internet]; 1993. Available from: http://books.google.com/books?hl=en&lr=&id=8y-FVtrKeSYC&oi=fnd&pg=PR7&dq=Against+method&ots=vC_G63Ox5D&sig=m7uNIrFo0VCGLTUVjk7vmnFSVm4

[2]   Kuhn, TS. *The structure of scientific revolutions.* In: Neurath O, editor. Philosophical review. London: University of Chicago Press; 1970. Vol. II, p. 210.

[3]   Lakatos, I. The methodology of scientific research programmes. Worrall J, Currie G, editors. 1st ed., Vol. 1. New York (NY): Press Syndicate of the University of Cambridge; 1978.

[4]   Arunachalam S. Open access to scientific knowledge. DESIDOC J Lib Inf Technol. 2008;28(1):7–14. Available from: http://www.publications.drdo.gov.in/ojs/index.php/djlit/article/view/147

[5] Molloy JC. The open knowledge foundation: open data means better science. PLoS Biol. 2011;9(12):e1001195. doi:10.1371/journal.pbio.1001195

[6] Crawford S, Stucki L. Peer review and the changing research record. J Am Soc Inf Sci. 1990;41(3):223–8. doi:10.1002/(SICI)1097-4571(199004)41:3%3C223::AID-ASI14%3E3.0.CO;2-3

[7] Pettifer S, McDermott P, Marsh J, Thorne D, Villeger A, Attwood TK. Ceci n'est pas un hamburger: modelling and representing the scholarly article. Learn Publ. 2011;24(3):207–20. doi:10.1087/20110309

[8] Houghton JW, Oppenheim C. The economic implications of alternative publishing models. Prometheus Crit Stud Innov. 2010;28(1):41–54. doi:10.1080/08109021003676359

[9] The Economist. Academic publishing: Free-for-all. The Economist [Internet]. 2013 [cited 18 May 2014]. Available from: http://www.economist.com/news/science-and-technology/21577035-open-access-scientific-publishing-gaining-ground-free-all

[10] Vision, TJ. Open data and the social contract of scientific publishing. BioScience. 2010;60(5):330–1. doi:10.1525/bio.2010.60.5.2

[11] Harnad S, Brody T, Vallières F, Carr L, Hitchcock S, Gingras Y, Oppenheim C, Hajjem C, Hilf ER. The access/impact problem and the green and gold roads to open access: an update. Serials Rev. 2008;34(1):36–40. doi:10.1016/j.serrev.2007.12.005

[12] Parks RP. The Faustian grip of academic publishing. J Econ Method. 2002;9(3):317–35. doi:10.1080/135017802200001 5122

[13] Van Noorden R. Open access: the true cost of science publishing. Nature. 2013;495(7442):426–9. doi:10.1038/495426a

[14] Kingsley, D. The journal is dead, long live the journal. Horizon. 2007;15(4):211–21. doi:10.1108/10748120710836237

[15] Poynder R. The OA interviews: Michael Eisen, co-founder of the public library of science. Open and Shut? [Internet]. 2012 [cited 20 May 2014]. Available from: http://perma.cc/G898-UCVT

[16] Brembs B. Libraries are better than corporate publishers because… Björn Brembs Blogarchive [Internet]. 2013 [cited 20 May 2014]. Available from: http://perma.cc/E5NY-KG4M

[17] Taylor M. Off-topic: non-open academic publishing is dead. Sauropod Vertebra Picture of the Week [Internet]. 2009 [cited 20 May 2014]. Available from: http://perma.cc/S4K6-NV42

[18] BOAI (Budapest Open Access Initiative). Budapest Open Access Initiative [Internet]; 2002 [cited 27 May 2013]. Available from: http://perma.cc/8BT3-KFNY

[19] Brown PO, Cabell D, Chakravarti A, Cohen B, Delamothe T, Eisen M, Grivell L, Guédon J-C, Hawley RS, Hawley RS, et al. Bethesda statement on open access publishing. Harvard Dash [Internet]. 2003 [cited 29 May 2013]. Available from: http://dash.harvard.edu/handle/1/4725199

[20] Max Planck Society. Berlin declaration. Open Access at the Max Planck Society; 2003. doi:10.1353/hrq.2005.0002

[21] Graziotin D, Wang X, Abrahamsson, P. A framework for systematic analysis of open access journals and its application in software engineering and information systems. Scientometrics. 2014;1–37. Digital Libraries; Computers and Society; Software Engineering. arXiv:1308.2597 [cs.DL]. doi:10.1007/s11192-014-1278-7

[22] IEEE. IEEE open – article processing charges. Institute of Electrical and Electornics Engineers [Internet]. 2013 [cited 20 May 2013]. Available from: http://perma.cc/LZ56-ARGG

[23] Bailey C. Open access and libraries. Collect Manage. 2008;32(3):351–83. doi:10.1300/J105v32n03_07

[24] Koehler, W. Web page change and persistence? A four-year longitudinal study. J Am Soc Inf Sci Technol. 2002;53(2):162–71. doi:10.1002/asi.10018

[24] Kingsley D. Walking in quicksand – keeping up with copyright agreements. Australian Open Access Support Group [Internet]. 2013 [cited 5 June 2014]. Available from: http://perma.cc/A5ZK-J928

[26] Harnad S. Gold open access publishing must not be allowed to retard the progress of green open access self-archiving. Logos. 2010;21(3):86–93. doi:10.1163/095796511X559972

[27] Xia J, Sun L. Assessment of self-archiving in institutional repositories: depositorship and full-text availability. Serials Rev. 2007;33(1):14–21. doi:10.1016/j.serrev.2006.12.003

[28] Swan A, Brown S. Open access self-archiving: an author study. Cornwall (UK); 2005. Vol. 44, p. 1–104. Available from: http://cogprints.org/4385

[29] Lawrence, S. Free online availability substantially increases a paper's impact. Nature. 2001;411(6837):521. doi:10.1038/35079151

[30] Davis PM, Fromerth, MJ. Does the arXiv lead to higher citations and reduced publisher downloads for mathematics articles? Scientometrics. 2007;71(2):203–15. doi:10.1007/s11192-007-1661-8

[31] Hajjem C, Harnad S, Gingras, Y. Ten-year cross-disciplinary comparison of the growth of open access and how it increases research citation impact. IEEE Data Eng Bull. 2005;28(4):39–47. Available from: http://eprints.ecs.soton.ac.uk/11688/2/Article IEEE.doc

[32] Kurtz MJ, Eichhorn G, Accomazzi A, Grant C, Demleitner M, Henneken E, Murray SS. The effect of use and access on citations. Inf Process Manage. 2005;41(6):1395–402. doi:10.1016/j.ipm.2005.03.010

[33] Craig ID, Plume AM, McVeigh ME, Pringle J, Amin, M. Do open access articles have greater citation impact? A critical review of the literature. J Informetrics. 2007;1(3):239–48. doi:10.1016/j.joi.2007.04.001

[34] Miller RM. Readers' attitudes to self-archiving in the UK. Napier University; 2006 Nov 20. [cited 18 May 2014]. Available from: http://eprints.rclis.org/8413/1/Rhiannon_Miller_self-archiving 2006.pdf

[35] Fronza I, Janes A, Sillitti A, Succi G, Trebeschi S. Cooperation wordle using pre-attentive processing techniques. In: 2013 6th International Workshop on Cooperative and Human Aspects of Software Engineering (CHASE). San Francisco (CA): IEEE; 2013. p. 57–64. doi:10.1109/CHASE.2013.6614732

[36] Björk B-C, Welling P, Laakso M, Majlender P, Hedlund T, Gudnason G. Open access to the scientific journal literature: situation 2009. PLoS One. 2010;5(6):e11273. doi:10.1371/journal.pone.0011273

[37] Gargouri Y, Larivière V, Gingras Y, Carr L, Harnad S. Green and gold open access percentages and growth, by discipline. In: 13th IEEE International Conference on Requirements Engineering (RE'05) 7th International Conference on Science and Technology Indicators. Montréal: STI; 2012; p. 285–92. Available from: http://arxiv.org/abs/1206.3664

[38] Howard J. Posting your latest article? You might have to take it down. The Chronicle of Higher Education [Internet]. 2013 [cited 18 May 2014]. Available from: http://perma.cc/Z6BE-JL6G

[39] Taylor M. Elsevier is taking down papers from Academia.edu. Sauropod Vertebra Picture of the Week [Internet]. 2013 [cited 18 May 2014]. Available from: http://perma.cc/GEL3-MXPR

[40] Ernesto. Publisher targets university researchers for "pirating" their own articles. TorrentFreak [Internet]. 2014 [cited 20 May 2014]. Available from: http://perma.cc/3MF6-JTFH

[41] Topsy.com. #ElsevierGate | Topsy.com. Topsy [Internet]. 2014 [cited 18 May 2014]. Available from: http://topsy.com/s?q= #ElsevierGate&window=a&type=tweet&language=en

**Competing Interests**

The author declares no competing interests.

**Publishing Notes**

Please note that this article may not have been peer reviewed yet and is under continuous post-publication peer review. For the current reviewing status please click **here** or scan the QR code on the right.

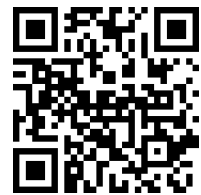